\documentclass[twocolumn,showpacs,preprintnumbers,amsmath,amssymb]{revtex4-1}

\usepackage{afterpage}
\usepackage[dvipdfmx]{graphicx}
\usepackage{color,bm}

\begin{document}

\title{
Gravitational lens on de-Sitter background
} 
\author{Keita Takizawa}
\author{Hideki Asada} 
\affiliation{
Graduate School of Science and Technology, Hirosaki University,
Aomori 036-8561, Japan} 
\date{\today}

\begin{abstract} 
Gravitational lenses are examined in de-Sitter (dS) background, 
for which the existence of the dS horizon is taken into account and 
hyperbolic trigonometry is used together with 
the hyperbolic angular diameter distance. 
Spherical trigonometry is used to discuss 
a gravitational lens in anti-de Sitter (AdS) background. 
The difference in the form among the dS/AdS lens equations 
and the exact lens equation in Minkowski background 
begins at the third order, 
when a small angle approximation is used 
in terms of lens and source planes. 
The angular separation of lensed images is 
decreased by the third-order deviation in the dS lens equation, 
while it is increased in AdS. 
In the present framework on the dS/AdS backgrounds, 
we discuss also the deflection angle of light, 
which does not include any term of purely the cosmological constant.  
Despite the different geometry, 
the deflection angle of light rays 
in hyperbolic and spherical geometry 
can take the same form. 
Through a coupling of the cosmological constant 
with lens mass, 
the separation angle of multiple images 
is larger (smaller) in dS (AdS) than in the flat case, 
for a given mass, source direction, and angular diameter distances 
among the lens, receiver and source.

\end{abstract}


\maketitle

\section{Introduction}
Since the observation by Eddington and his collaborators 
\cite{Eddington}, 
the gravitational deflection of light has played an important role 
in astronomy and gravitational physics. 
The Event Horizon Telescope (EHT) team has recently succeeded 
a direct image of the immediate vicinity of the supermassive black hole (BH) 
candidate of M87 galaxy \cite{EHT}. 
In addition, the same team has reported measurements 
of linear polarizations around the BH candidate
\cite{EHT2021a} 
and has inferred the mass accretion rate and 
the strength of the magnetic field \cite{EHT2021b}. 
Such groundbreaking events have generated 
renewed interest in the strong field gravitational lens.

\begin{figure}
\includegraphics[width=8.6cm]{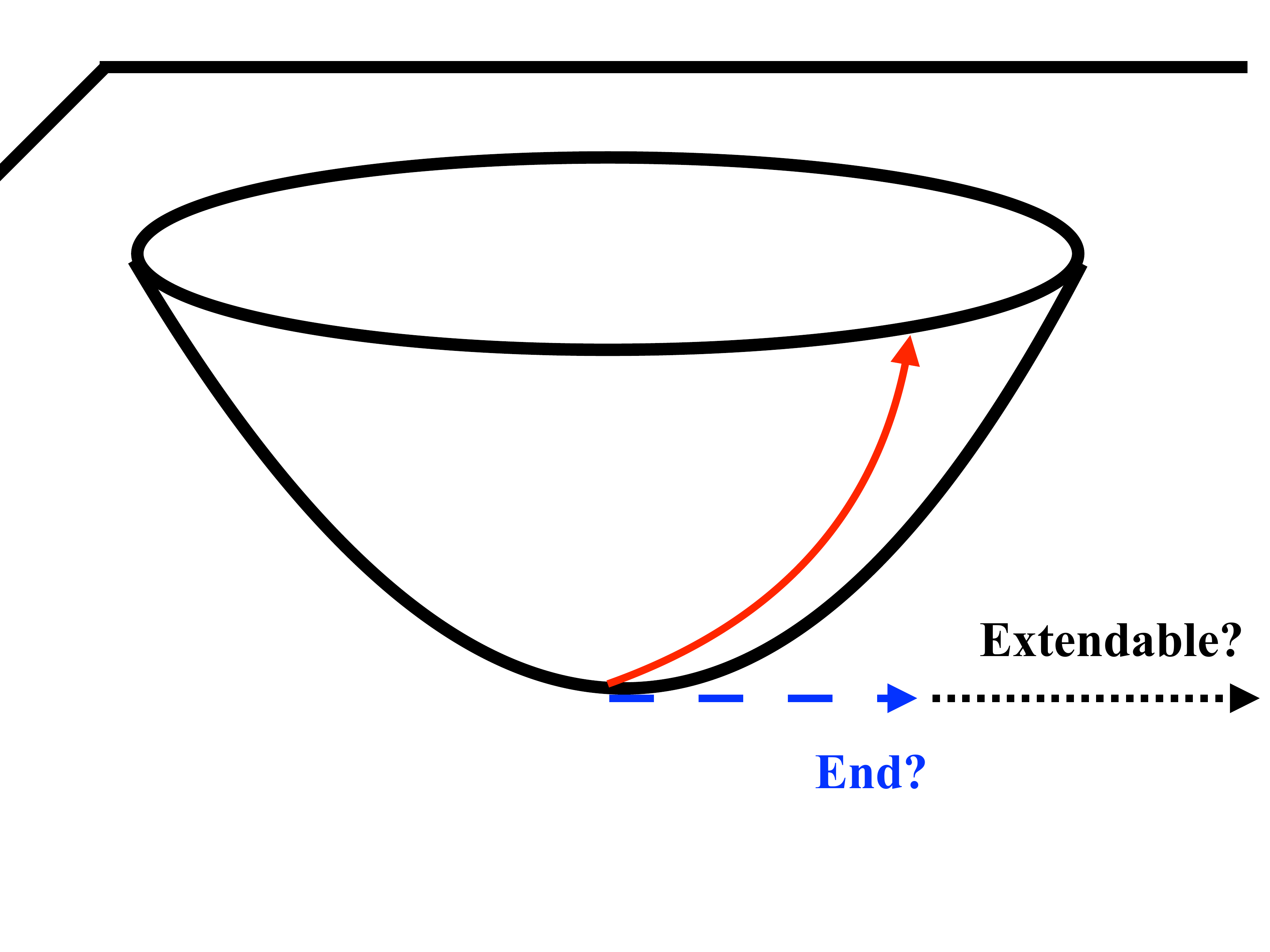}
\caption{
Intuitive illustration of 
how an orbit in a curved surface with a boundary 
and its approximation on a flat background are different. 
The solid red (in color) arrow denotes a curve on a bowl 
and the dashed blue (in color) arrow denotes a line on a table. 
On the flat background, the dashed arrow 
means an approximation of the curve on the bowl. 
The approximation is good at small scale 
in the neighborhood of the bottom of the bowl, 
while the difference of the two curves becomes significant 
at very large scale, especially near the edge of the bowl. 
The edge has an analogy with 
the so-called de-Sitter horizon in de-Sitter spacetime, 
though the latter is traversable one way. 
The solid arrow terminates at the edge of the bowl. 
On the other hand, one may ask if the dashed line 
can be more extended as indicated by the dotted black (in color) arrow. 
}
\label{fig-idea}
\end{figure}

Gravitational lens equations, 
which relate source and image positions 
for a given deflecting mass, 
are at the heart of the gravitational lens theory. 
For most lens equations, 
the background spacetime is assumed to be Minkowskian. 
In the flat background, Euclidean geometry holds for its spatial sector. 
Therefore, Euclidean trigonometry is used to arrive at 
the standard form of the lens equation \cite{SEF, Petters, Dodelson, Keeton}. 

However, the current and future astronomy needs a more precise prediction 
from a gravitational lensing theory especially relevant with the strong field regime. 
The Schwarzschild de-Sitter spacetime (often called Kottler solution) 
plays a theoretical model that allows us to examine the gravitational deflection 
of light by a mass in the presence of the cosmological constant $\Lambda$. 
Many attempts have been made on the simple model 
in the context of the gravitational lensing e.g. 
\cite{Lake,RI,Finelli,Park,Sereno,Bhadra,Simpson,AK,Ishihara2016,Piattella,Butcher,LW,AB,Arakida2021,LLJ}. 
Intuitively, the Minkowskian background can work as an approximation at small scale, 
though there can be a significant departure of the Minkowskian background 
from de-Sitter backgrounds especially at very large scale. 
See Figure \ref{fig-idea} for this intuition. 
Until recently, the definition for the deflection angle of light has required 
the asymptotic flatness of a spacetime 
\cite{SEF, Petters, Dodelson, Keeton}. 
However, recent works 
\cite{Ishihara2016, Takizawa2020a} 
based on the Gauss-Bonnet theorem 
\cite{GW} 
have enabled us to well define the deflection angle 
without assuming the asymptotic flatness. 
This new method has been applied by several groups to various spacetime models 
\cite{Ishihara2017, Jusufi2017a, Jusufi2017b, Jusufi2018, Ovgun2019, Ovgun2020, Crisnejo2018, Crisnejo2019, Ono2017, Ono2018, Ono2019}. 
See Reference \cite{Ono2019b} for a brief review on this subject.

Can the lens equation on de-Sitter backgrounds 
take the same form 
as that in Minkowskian background? 
One may specifically ask if  
the de-Sitter lens equation can be expressed in the same form 
when angular diameter distances are properly defined. 
One possible answer is that there would exist 
a difference between the two lens equations 
in the flat and de-Sitter backgrounds, 
even if angular diameter distances are defined in a suitable manner. 
If this answer is correct, 
what changes are made? 

The main purpose of this paper is to examine gravitational lenses 
on de-Sitter background. 
Henceforth, cases of $\Lambda > 0$ and $\Lambda < 0$ 
are referred to as dS and AdS, respectively. 
A key tool in the present study is the optical metric, which is known to 
describe light rays in stationary spacetimes 
\cite{AK2000, GW, Ishihara2016,Ono2017}. 
We shall show that the optical metric for dS spacetime describes 
a hyperbolic space, 
while that for AdS spacetime corresponds to 
spherical geometry. 
Therefore, we use hyperbolic trigonometry 
in order to derive a gravitational lens equation on the dS background, 
where we do not use a small angle approximation nor 
a thin lens one. 
For AdS case, 
we use spherical trigonometry to 
derive a gravitational lens equation on the AdS background. 
In both the derivations, 
we define angular diameter distances on dS/AdS backgrounds. 

This paper is organized as follows. 
In Section II, 
the optical metric is examined on dS/AdS backgrounds. 
By using hyperbolic trigonometry, 
in Section III, 
we discuss gravitational lens equations 
on dS background. 
In Section IV, 
we make use of spherical trigonometry 
to study gravitational lens equations 
on AdS background. 
In Section V, 
the Euclidean, hyperbolic and spherical lens equations 
are unified into a single equation. 
The deflection angle of light on dS/AdS backgrounds 
also is examined. 
Finally, this paper is summarized in Section VI. 
Throughout this paper, we use the unit of $G = c = 1$.

\section{Optical metric for dS/AdS spacetimes}
\subsection{Hyperbolic space:  dS case}
There are several known ways for slicing of dS spacetime, 
where the scalar curvature of dS universe is positive. 
We work on the static coordinates of dS spacetime, 
where the dS metric reads 
\begin{align}
  d s^2 =& -\left(1 - \frac{\Lambda}{3}r^2\right) d t^2 
 + \cfrac{d r^2}{1 - \cfrac{\Lambda}{3}r^2}
+ r^2( d\Theta^2 + \sin^2 \Theta d\phi^2) ,
\label{dS}
\end{align}
where $r_H \equiv \sqrt{3/\Lambda}$ is a radius of the dS horizon.

For later convenience, we normalize a radial coordinate 
in terms of the cosmological constant as 
\begin{equation}
R \equiv \sqrt{\frac{\Lambda}{3}} r , 
\label{R} 
\end{equation}
such that dS metric can be reexpressed as 
\begin{align}
  d s^2 =& -\left(1 - R^2\right) d t^2 
  \nonumber\\
  &+ 
 \frac{3}{\Lambda} 
 \left(
 \cfrac{d R^2}{1 - R^2}
+ R^2( d\Theta^2 + \sin^2 \Theta d\phi^2) 
\right) .
\label{dS-2}
\end{align}
In the present paper, 
we focus on a case that the lens, receiver and source are located 
inside the dS horizon, namely $R < 1$.

We consider the optical metric followed by the null condition $ds^2 = 0$, 
which determines the light propagation 
\cite{AK2000, GW, Ishihara2016} 
\begin{align}
d\ell^2 &\equiv dt^2 
\nonumber\\
&= 
\frac{3}{\Lambda} 
 \left(
 \cfrac{d R^2}{(1 - R^2)^2}
+ \frac{R^2}{1 - R^2}( d\Theta^2 + \sin^2 \Theta d\phi^2) 
\right) . 
\label{optical}
\end{align}

Clearly, it is convenient to work on the conformally rescaled metric as 
\begin{align}
d\hat\ell^2 &\equiv \frac{\Lambda}{3} d\ell^2 
\nonumber\\
&= 
 \cfrac{d R^2}{(1 - R^2)^2}
+ \frac{R^2}{1 - R^2}( d\Theta^2 + \sin^2 \Theta d\phi^2) . 
\label{optical-2}
\end{align}

We define a new radial coordinate $\rho$ by 
\cite{footnote-1}
\begin{align}
R \equiv \tanh \rho , 
\label{rho}
\end{align}
to reexpress the rescaled metric as 
\begin{align}
d\hat\ell^2 
&= 
 d\rho^2
+ \sinh^2\rho( d\Theta^2 + \sin^2 \Theta d\phi^2) . 
\label{optical-3}
\end{align}

Note that $\rho$ coordinate is different from $\chi$ coordinate 
that is usually used in general relativistic cosmology. 
Eq. (\ref{optical-3}) means that the optical metric in dS case describes 
a hyperbolic space, in which unlensed light rays are geodesic curves. 

The dS horizon is located at $\rho = +\infty$ 
in the metric Eq. (\ref{optical-3}). 
This means that $\rho$ coordinate describes only the inside 
of the dS horizon. 
On the other hand, it is apparent that the $r$ coordinate 
can go beyond the dS horizon, if one treats perturbatively Eq. (\ref{dS}) 
around the Minkowskian background even at very large scale.

\subsection{Spherical space: AdS case}
For AdS case ($\Lambda < 0$), 
all we have to do is 
to replace $\sin\chi$ by $\sinh\chi$. 
For this, we introduce 
\cite{footnote-2}
\begin{align}
R &\equiv \sqrt{\frac{-\Lambda}{3}} r ,
\\
R&\equiv \tan\rho ,
\end{align}
to obtain the normalized optical metric as 
\begin{align}
d\hat\ell^2 
&= 
 d\rho^2
+ \sin^2\rho( d\Theta^2 + \sin^2 \Theta d\phi^2) . 
\label{optical-AdS}
\end{align}
This is a metric of spherical geometry.

\section{Gravitational lens in hyperbolic geometry}
\subsection{Hyperbolic trigonometry}
We consider a photon orbit around a spherically symmetric lens 
at the origin of the spatial coordinates $\rho=0$. 
Because of the symmetry, we can choose the equatorial plane $\Theta = \pi/2$ 
as a photon orbit for its simplicity. 
This choice can be done also for a photon orbit 
in the equatorial plane of an axisymmetric lens 
with reflection symmetry. 

Then, we can use 
the conformally normalized optical metric of a hyperbolic plane as 
\begin{align}
d\hat\ell^2 
&= 
 d\rho^2
+ \sinh^2\rho d\phi^2 . 
\label{optical-equatorial}
\end{align}
It follows that the hyperbolic plane is homogeneous and isotropic. 
This allows us to explicitly define angular diameter distance 
in a simple manner. 
See Section III for more detail.

\begin{figure}
\includegraphics[width=8.6cm]{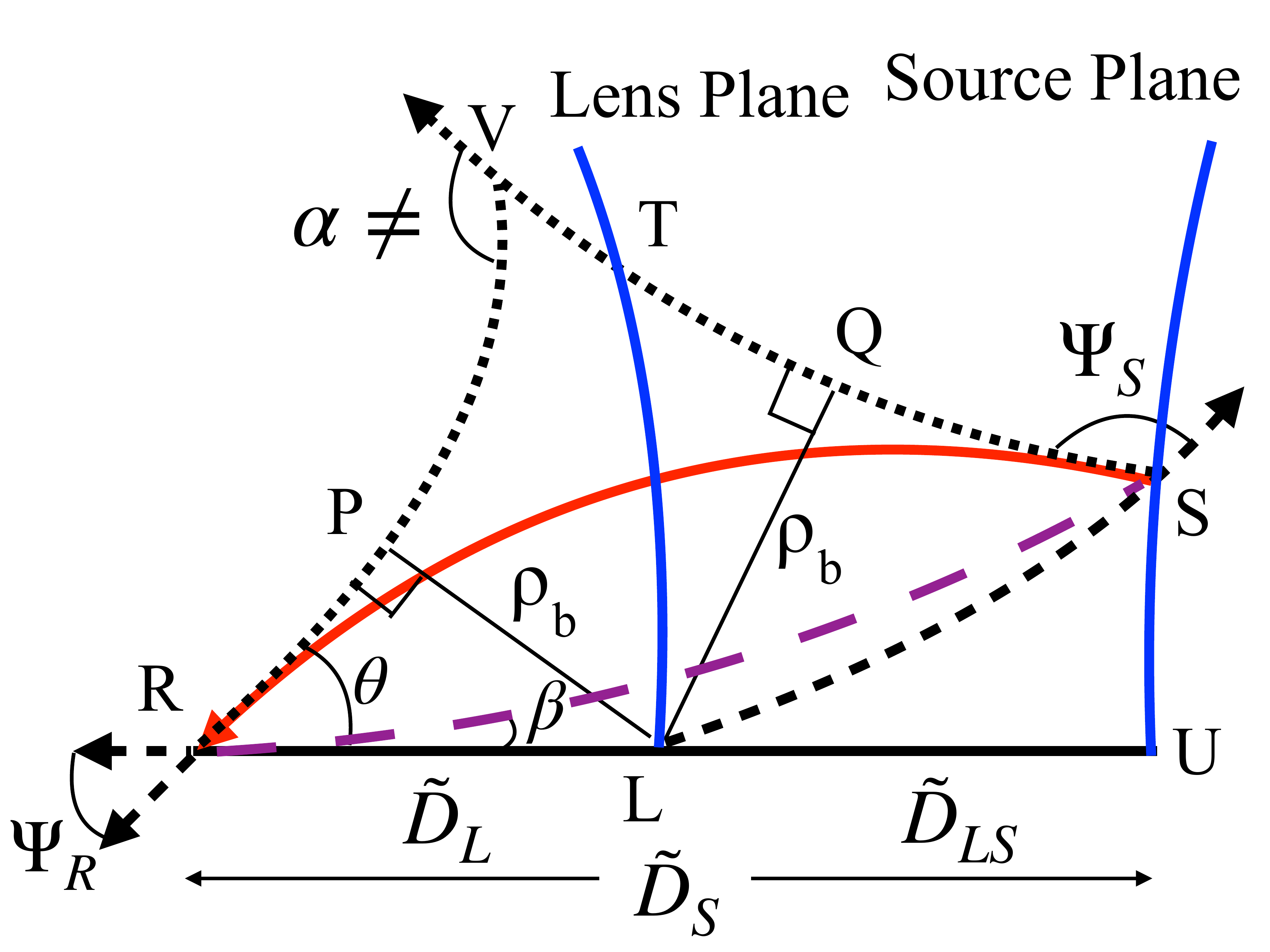}
\caption{
Schematic figure of a lens $L$, receiver $R$ and source $S$  
in a hyperbolic plane. 
For its simplicity, a line connecting L and R is drawn as a straight one 
for reference. 
The red (in color) curve connecting R and S denotes a {\it lensed} light ray, 
which is not a geodesic in the hyperbolic plane. 
A dotted geodesic curve emanating from S is a tangent 
to the lensed light ray, 
while the other dotted one from R is another tangent to the light ray. 
The latter dotted line indicates the lensed image direction 
$\theta (= \Psi_R)$ 
seen from the receiver. 
A geodesic curve between R and S is denoted by a long dashed line, 
which indicates the unlensed source direction $\beta$. 
The lens and source planes are vertical to the geodesic line RU. 
These curves are not parallel to each other 
because of hyperbolicity. 
Note that the sum of the inner angles for the hyperbolic quadrilateral 
LRVS does not equal to $2\pi$ according to the Gauss-Bonnet theorem 
for the curved surface (See e.g. \cite{Ishihara2016, Takizawa2020b}). 
Therefore, 
the outer angle at the intersection point V of the two tangents 
in the hyperbolic plane 
differs from 
$\alpha (= \Psi_R - \Psi_S + \phi_{RS})$. 
In fact, the intersection V has nothing to do with 
the derivation of the lens equations in this paper. 
}
\label{fig-config-h}
\end{figure}

\begin{figure}
\includegraphics[width=8.6cm]{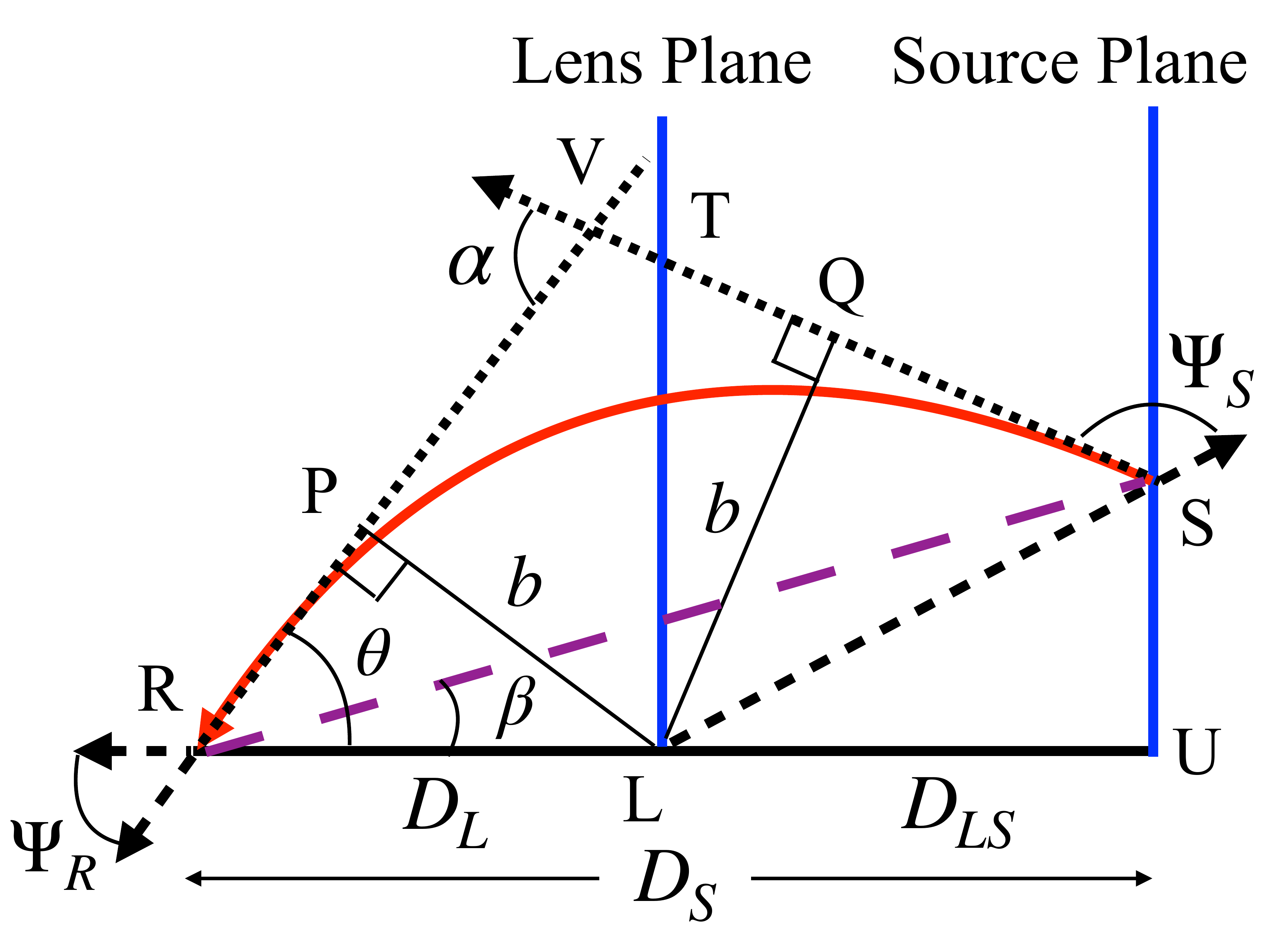}
\caption{
A gravitational lens system in a Euclidean background 
\cite{Takizawa2020b}. 
The notations are the same as those in Figure \ref{fig-config-h}.  
The lens and source planes are parallel to each other, 
because they live in a Euclidean space. 
}
\label{fig-config-f}
\end{figure}

We consider hyperbolic triangles 
in order to discuss a gravitational lens configuration 
on a hyperbolic plane. 
Figure \ref{fig-config-h} is a schematic illustration 
of the gravitational lens configuration in the hyperbolic plane. 
For a comparison, see also Figure \ref{fig-config-f} 
for a flat background.

\begin{figure}
\includegraphics[width=8.6cm]{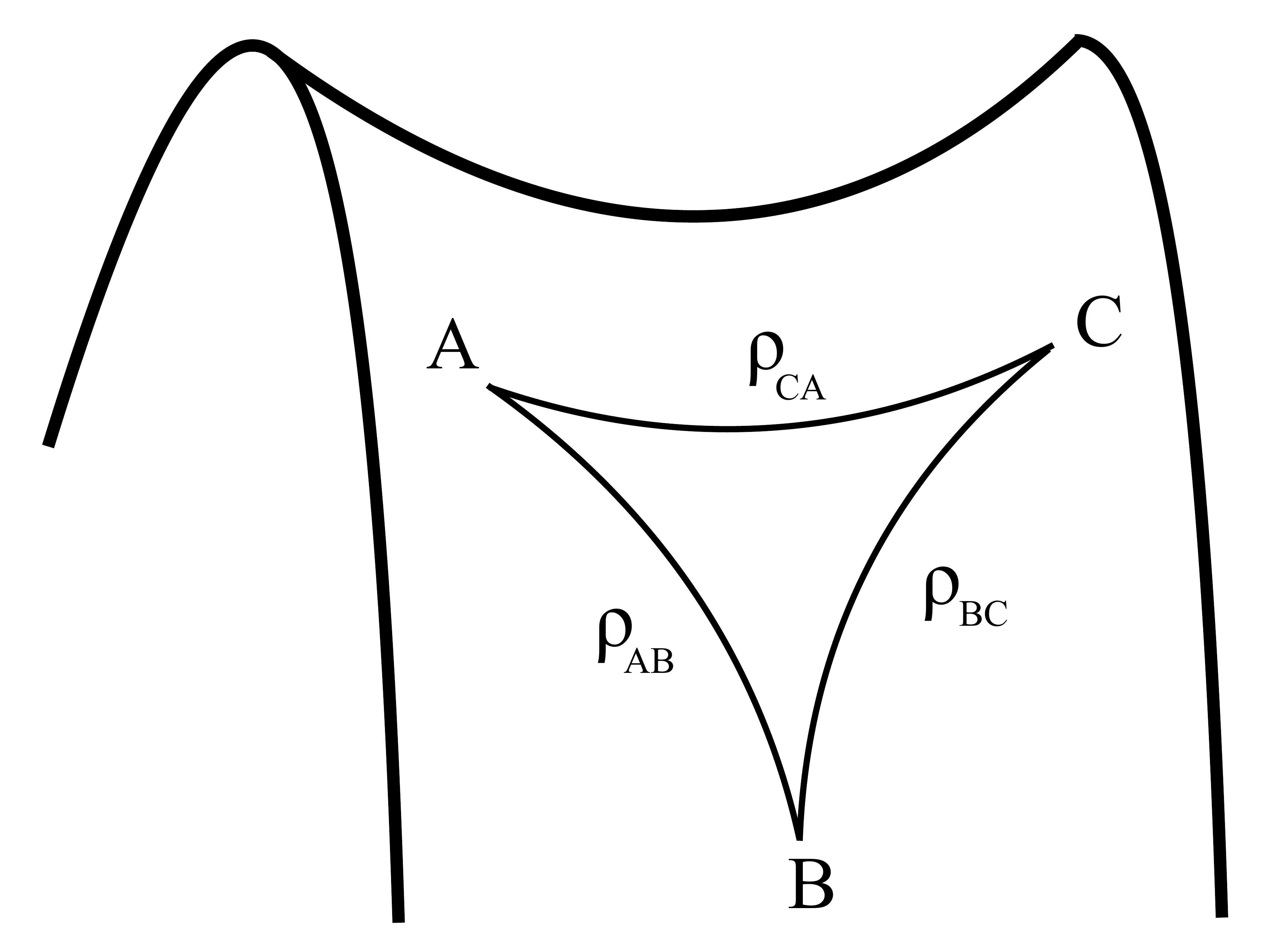}
\caption{
Schematic figure of a hyperbolic triangle ABC. 
The vertices are denoted by A, B and C. 
The sides of the triangle are geodesics 
indicated by a thin solid line. 
The arc length of them is   
$\rho_{AB}$, $\rho_{BC}$ and $\rho_{CA}$. 
}
\label{fig-triangle-h}
\end{figure}

First, we briefly summarize hyperbolic trigonometry 
\cite{Anderson-book, Ratcliffe-book}, 
with which most physicists may not be familiar. 
Let us consider a triangle $ABC$, where $A$, $B$ and $C$ 
denote the vertices of the triangle and they mean also the inner angles. 
The sides of the triangle follow geodesics in the hyperbolic plane. 
See Figure \ref{fig-triangle-h} for the hyperbolic triangle $ABC$. 

In all the formulae stated here,  
arc length of the sides $a$, $b$ and $c$ are measured in absolute unit, 
in which the Gaussian curvature $K$ of the plane is $-1$. 
For instance, in this paper, the absolute length of the side $AB$ 
(which is in the normalized optical metric) is 
denoted as $\rho_{AB}$, 
which is different from the proper length measured 
by the original optical metric. 
In the case that $A$ is the receiver, 
we shall omit it simply as 
$\rho_B$, 
because the receiver is often chosen as the coordinate origin 
in the gravitational lens study.

The hyperbolic law of sines is 
\begin{align}
\frac{\sinh\rho_{BC}}{\sin A} 
= 
\frac{\sinh\rho_{CA}}{\sin B} 
=
\frac{\sinh\rho_{AB}}{\sin C} . 
\label{sine-h}
\end{align}
The hyperbolic law of cosines is 
\begin{align}
\cosh\rho_{AB} = \cosh\rho_{BC} \cosh\rho_{CA} - \sinh\rho_{BC} \sinh\rho_{CA} \cos C . 
\label{cos-h}
\end{align}

The trigonometry of right angles is as follows. 
If $C = \pi/2$, then 
\begin{align}
\sin A =& 
\frac{\sinh\rho_{BC}}{\sinh\rho_{AB}} , 
\label{sin-h-R}
\\
\cos A =&
\frac{\cosh\rho_{BC}\sinh\rho_{CA}}{\sinh \rho_{AB}}
\notag\\
=&
\frac{\tanh\rho_{CA}}{\tanh\rho_{AB}} , 
\label{cos-h-R}
\\
\cosh\rho_{AB} =&
 \cosh\rho_{BC} \cosh\rho_{CA} . 
 \label{cosh-R}
\end{align} 
Eq. (\ref{cosh-R}) is a hyperbolic generalization of Pythagorean theorem. 
It recovers $\rho_{AB}^2 = \rho_{BC}^2 + \rho_{CA}^2$ 
at very small scale of $\rho_{AB} \ll 1$.

\subsection{Lens, receiver and source in a hyperbolic plane}
For the right triangle PRL, we obtain 
\begin{align}
\sin\theta 
=&
\frac{\sinh \rho_{PL}}{\sinh \rho_L} ,
\label{PRL-h}
\end{align}
where Eq. (\ref{sin-h-R}) is used. 
Similarly, for the right triangle LQS, 
\begin{align}
\sin(\pi - \Psi_S)  
=&
\frac{\sinh \rho_{QL}}{\sinh \rho_{LS}} .
\label{LQS-h}
\end{align}

Both of $\rho_{PL}$ and $\rho_{QL}$ 
mean the arc length $\rho_b$ of the impact parameter 
of a single light ray in the normalized optical metric. 
See Figure \ref{fig-config-h}. 
For the single light ray, they must be equal to each other. 
Therefore, $\rho_{PL} = \rho_{QL}$. 
By eliminating $\rho_{PL}$ and $\rho_{QL}$ 
from Eqs. (\ref{PRL-h}) and (\ref{LQS-h}), 
we thus find a relation between $\Psi_S$ and $\theta$ as 
\begin{align}
\Psi_S 
= 
\pi - \arcsin\left(\frac{\sinh\rho_L}{\sinh\rho_{LS}}\sin\theta\right) .
\label{PsiS-h}
\end{align}

Next, we consider the right triangle LSU, 
where the point U is chosen 
such that the side SU is perpendicular to LU. 
In the hyperbolic plane, the inner angle between LR and LS 
is the same as the longitude $\phi_{RS}$ 
in dS spacetime. 
Using Eq. (\ref{sin-h-R}) at the vertex L of LSU, 
we find 
\begin{align}
\sin(\pi - \phi_{RS}) 
=
\frac{\sinh\rho_{SU}}{\sinh\rho_{LS}} . 
\label{LSU-h}
\end{align}

Similarly, from the triangle RSU, 
we obtain 
\begin{align}
\sin\beta 
=
\frac{\sinh\rho_{SU}}{\sinh\rho_S} . 
\label{RSU-h}
\end{align}
Here, 
$\beta$ denotes the angle $R$, which 
is the same as the directional angle of the unlensed source in dS spacetime. 
From Eqs. (\ref{LSU-h}) and (\ref{RSU-h}), 
we obtain a relation 
between the longitude $\phi_{RS}$ and the source direction $\beta$ as 
\begin{align}
\phi_{RS} 
= 
\pi - \arcsin\left(\frac{\sinh\rho_S}{\sinh\rho_{LS}}\sin\beta\right) .
\label{phiRS-h}
\end{align}

We work on 
the angle of the gravitational deflection of light 
that has been defined by Ishihara et al. as \cite{Ishihara2016} 
\begin{align}
\alpha \equiv \theta -\Psi_S + \phi_{RS} .
\label{alpha}
\end{align}
It  has been recently proven that 
the definition holds even in a non-asymptotically flat spacetime 
\cite{Takizawa2020b}. 

By substituting Eqs. (\ref{PsiS-h}) and (\ref{phiRS-h}) into Eq. (\ref{alpha}), 
we obtain 
\begin{align}
\alpha - \theta 
= 
\arcsin\left(\frac{\sinh\rho_L}{\sinh\rho_{LS}}\sin\theta\right)
- \arcsin\left(\frac{\sinh\rho_S}{\sinh\rho_{LS}}\sin\beta\right) . 
\label{lenseq-dS}
\end{align}

\subsection{Hyperbolic angular diameter distance}
In the normalized hyperbolic plane, 
$\hat d$ denotes the angular diameter distance 
between two points, 
while we use the conventional notation $d$ for 
the angular diameter distance in the original metric. 
In dS case, the physical angular diameter distance is 
$d = \hat d \sqrt{3/\Lambda}$. 
From Eq. (\ref{optical-equatorial}), 
the angular diameter distance is defined as 
\begin{align}
\hat d_L 
&\equiv \sinh\rho_L , 
\label{dL-h}
\\
\hat d_S 
&\equiv \sinh\rho_S , 
\label{dS-h}
\\
\hat d_{LS} 
&\equiv \sinh\rho_{LS} . 
\label{dLS-h}
\end{align}
Thereby, Eq. (\ref{lenseq-dS}) can be rewritten explicitly 
in terms of the angular diameter distance as 
\begin{align}
\alpha - \theta 
= 
\arcsin\left(\frac{\hat d_L}{\hat d_{LS}}\sin\theta\right)
- \arcsin\left(\frac{\hat d_S}{\hat d_{LS}}\sin\beta\right) . 
\label{lenseq-h-d}
\end{align}

Eq. (\ref{lenseq-h-d}) takes the same form as 
Eq. (17) of Reference \cite{Takizawa2020b} 
for the flat background 
\cite{footnote-3}. 
It is not a surprising coincidence, 
because the both derivations do not rely 
on whether the quadrilateral LRQS lives in a flat space. 
The sum of the internal angles is not necessarily 
$L+R+V+S = 2\pi$. 
Indeed, the formulation by Takizawa et al. 
\cite{Takizawa2020b} stands on fully curved backgrounds. 
However, this point has not been stressed 
\cite{Takizawa2020b}. 

We should note also that the addition law holds 
for the flat angular diameter distance 
in a Euclidean space, 
but it does not in a curved space. 
Indeed, $\hat d_L + \hat d_{LS} \neq \hat d_S$ in hyperbolic geometry.

\subsection{Lens and source planes in hyperbolic geometry}
In the above, we have considered the angular diameter distance between points. 
On the other hand, most studies on the gravitational lens employ 
angular diameter distances between a point (usually chosen as the receiver position) 
and the so-called lens plane (or the source plane). 
In a Euclidean space, 
the lens and source planes are parallel to each other. 
In a non-Euclidean space, however, the two planes are not parallel to each other.

In the hyperbolic space, we define a lens plane as a surface which consists of 
a family of geodesics, every of which is vertical to the geodetic line LR 
at the point L. 
We consider a point in the hyperbolic space, 
at which the geodetic line connecting L and R is vertical 
to another geodetic emanating from the source point S. 
We denote the point as U. 
By the same way, we define a source plane that is vertical to 
the geodetic line connecting L and R at the point U.  
See Figure \ref{fig-config-h} for the lens and source planes. 

In terms of the lens and source planes, 
we can define the angular diameter distances. 
The normalized angular diameter distance from the receiver to the lens plane 
is defined as that from the receiver to the point L on the lens plane. 
That is,  
\begin{align}
\hat D_L 
&\equiv 
\hat d_L 
\notag\\
&= \sinh\rho_L , 
\label{DL-h}
\end{align}
where we use Eq. (\ref{dL-h}). 

In the similar manner, 
the normalized angular distance from the receiver to the hyperbolic source plane 
is defined as that from the receiver to the point U on the source plane 
\begin{align}
\hat D_S 
\equiv 
\sinh\rho_U . 
\label{DS-h}
\end{align} 
The normalized angular distance between the lens and source planes  
is defined as that from the point L to the point U 
\begin{align}
\hat D_{LS} 
&\equiv 
\sinh\rho_{LU} .
\label{DLS-h}
\end{align}
Note that $\hat D_L + \hat D_{LS} \neq \hat D_S$ in hyperbolic geometry. 
This is because 
$\sinh\rho_L + \sinh\rho_{LU} \neq \sinh\rho_U$, 
though $\rho_L + \rho_{LU} = \rho_U$. 
Again, we should note that 
the physical angular diameter distance needs the factor as 
$D = \hat D \sqrt{3/\Lambda}$.

By using the cosine formula Eq. (\ref{cos-h-R}) 
for the right triangle RSU, 
we obtain 
\begin{align}
\cos\beta 
= 
\frac{\cosh \rho_{SU}\sinh\rho_U}{\sinh\rho_S} .
\label{cos-beta-h}
\end{align}
From Eqs. (\ref{RSU-h}) and (\ref{cos-beta-h}), 
we find 
\begin{align}
\tan\beta 
&=
\frac{\sin\beta}{\cos\beta}
\notag\\
&=
\frac{\tanh\rho_{SU}}{\sinh\rho_U} . 
\label{tanbeta-h}
\end{align}
By using this for Eq. (\ref{DS-h}), we obtain 
\begin{align}
\hat D_S 
= 
\frac{\tanh\rho_{SU}}{\tan\beta} .
\end{align}
From this, we can see 
\begin{align}
\sinh^2\rho_{SU} 
=
\frac{\hat D_S^2 \tan^2\beta}{1 - \hat D_S^2 \tan^2\beta} . 
\label{sinh2SU-h}
\end{align}

By using Eq. (\ref{cosh-R}) for the right triangle LSU, 
we obtain 
\begin{align}
\cosh\rho_{LS} = \cosh\rho_{SU} \cosh\rho_{LU} . 
\label{coshLS}
\end{align}
This leads to 
\begin{align}
\sinh^2\rho_{LS} 
&=
\cosh^2\rho_{LS} - 1 
\notag\\
&=
\left(\cosh\rho_{LU} \cosh\rho_{SU}\right)^2 - 1 
\notag\\
&=
\frac{\hat D_{LS}^2 + \hat D_S^2 \tan^2\beta}
{1 - \hat D_S^2 \tan^2\beta} ,
\label{sinh2LS}
\end{align}
where Eqs. (\ref{DLS-h}) and (\ref{sinh2SU-h}) are 
used in the last line.

Next, we consider the angle L in the right triangle LSU. 
We find 
\begin{align}
\sin L = \frac{\hat d_S}{\hat d_{LS}} \sin\beta ,
\label{sinL-h}
\end{align}
where Eq. (\ref{sin-h-R}) is used. 
We thus obtain  
\begin{align}
\tan L 
&= 
\frac{\sin L}{\cos L} 
\notag\\
&= 
\frac{\hat D_S}{\hat D_{LS}} \tan\beta , 
\label{tanL-h}
\end{align}
where Eqs. (\ref{cos-h-R}), (\ref{cosh-R}) and (\ref{sinh2LS}) are used. 
By combining Eqs. (\ref{sinL-h}) and (\ref{tanL-h}), 
we obtain
\begin{align}
\arcsin
\left(
\frac{\hat d_S}{\hat d_{LS}} \sin\beta 
\right)
=
\arctan
\left(
\frac{\hat D_S}{\hat D_{LS}} \tan\beta 
\right) .
\label{arcsintoarctan-h}
\end{align}

In terms of the angular diameter distance $\hat D$, 
Eq. (\ref{lenseq-h-d}) can be reexpressed as 
\begin{align}
\alpha - \theta 
=& 
\arcsin\left(
\sqrt{
\frac{1 - \hat D_S^2 \tan^2\beta}{\hat D_{LS}^2 + \hat D_S^2 \tan^2\beta}
}
\hat D_L\sin\theta\right) 
\notag\\
& 
- 
\arctan\left(\frac{\hat D_S}{\hat D_{LS}}\tan\beta\right) ,
\label{lenseq-h-D}
\end{align}
where Eqs. (\ref{DL-h}), (\ref{DS-h}), (\ref{DLS-h}), (\ref{tanbeta-h}) 
(\ref{sinh2LS}) and (\ref{arcsintoarctan-h}) are used. 
The inside of the square root in Eq. (\ref{lenseq-h-D}) must be nonnegative. 
Therefore, 
\begin{align}
|\tan\beta| \leq \frac{1}{\hat D_S} , 
\label{beta-bound}
\end{align}
which gives the allowed region of the source direction. 
This is due to the existence of the dS horizon.

The hyperbolic lens equation Eq. (\ref{lenseq-h-D}) 
is slightly different from the Takizawa lens equation in the flat background as 
\cite{Takizawa2020b} 
\begin{align}
\alpha - \theta 
=& 
\arcsin\left(
\frac{D_L}{\sqrt{D_{LS}^2 + D_S^2 \tan^2\beta}}
\sin\theta\right) 
\notag\\
& 
- 
\arctan\left(\frac{D_S}{D_{LS}}\tan\beta\right) .
\label{lenseq-f-D}
\end{align}

The only difference is caused because Euclidean Pythagorean theorem does not stand 
in hyperbolic space, especially for the right triangle such as RSU and LSU. 
In other words, 
the methods of deriving Eqs. (\ref{lenseq-h-D}) and (\ref{lenseq-f-D}) 
do depend on whether the triangle LSU lives in a flat space. 

Eqs. (\ref{lenseq-h-D}) and (\ref{lenseq-f-D}) bear a striking resemblance to each other, 
though they are based on two completely different geometry. 
Does it mean that the cosmological constant makes almost 
no effects on gravitational lens observations? 
No. 
Eq. (\ref{lenseq-h-D}) is written by using the angular diameter distances 
in hyperbolic geometry due to the presence of the cosmological constant. 
This means that the cosmological constant significantly affects 
the gravitational lens.

\section{Gravitational lens in spherical geometry} 
\subsection{Spherical trigonometry} 
In the similar manner to the previous section, 
we focus on photon orbits on the equatorial plane $\theta = \pi/2$ 
in spherical geometry, which corresponds to AdS case. 
Then, the normalized optical metric in the plane is 
\begin{align}
d\hat\ell^2 
&= 
 d\rho^2
+ \sin^2\rho d\phi^2 . 
\label{optical-equatorial-AdS}
\end{align}

\begin{figure}
\includegraphics[width=8.6cm]{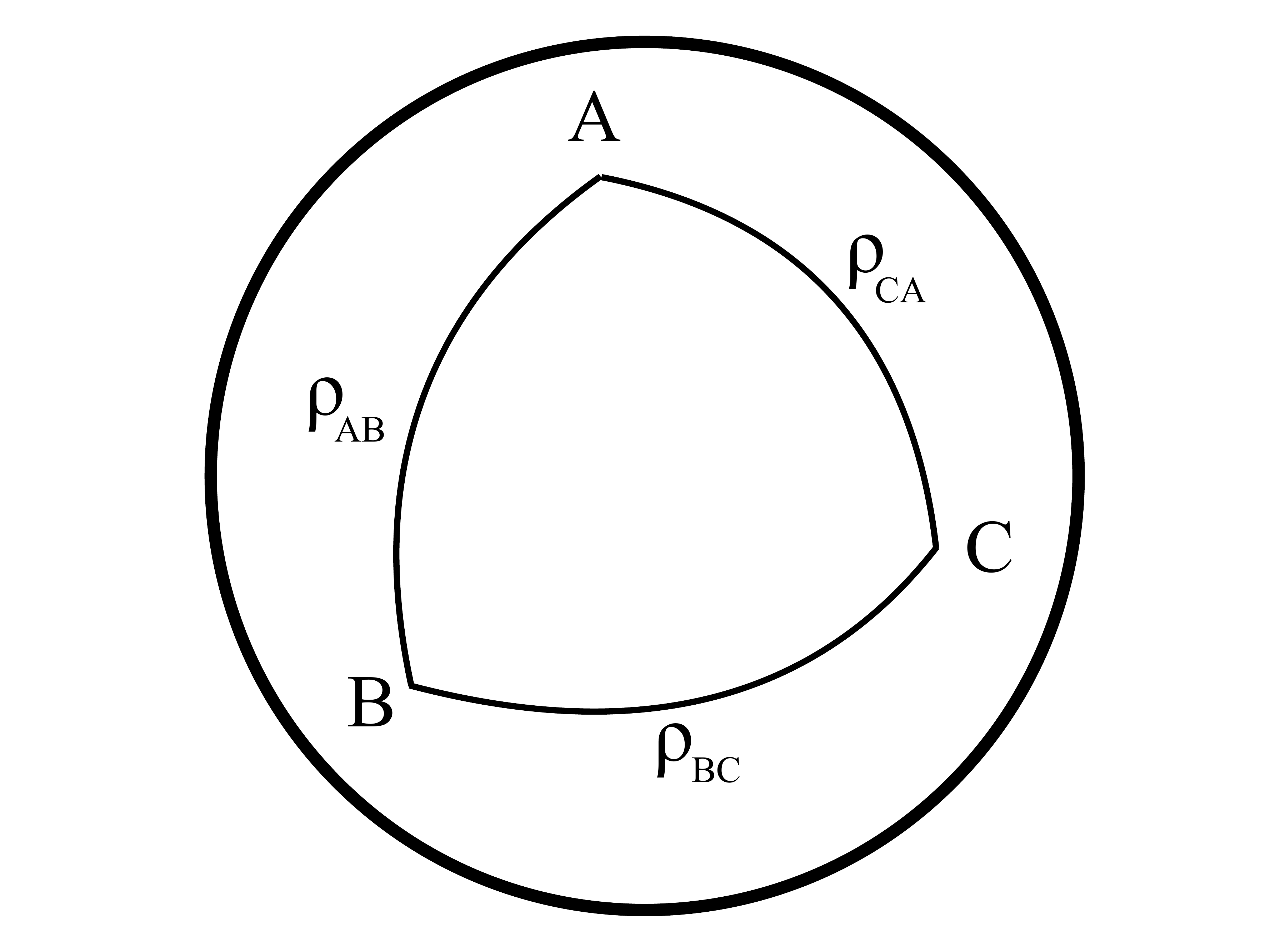}
\caption{
Schematic figure of a spherical triangle. 
The notations are the same as those in Figure \ref{fig-triangle-h}
}
\label{fig-triangle-s}
\end{figure}

Let us briefly summarize the spherical trigonometry 
\cite{Ratcliffe-book}. 
See Figure \ref{fig-triangle-s} for a spherical triangle $ABC$. 

The laws of sines and cosines are 
\begin{align}
\frac{\sin\rho_{BC}}{\sin A} 
= 
\frac{\sin\rho_{CA}}{\sin B} 
=
\frac{\sin\rho_{AB}}{\sin C} , 
\label{sine-s}
\end{align}
\begin{align}
\cos\rho_{AB} = \cos\rho_{BC} \cos\rho_{CA} - \sin\rho_{BC} \sin\rho_{CA} \cos C .
\label{cos-s}
\end{align}

The spherical trigonometry of right angles is as follows. 
If $C = \pi/2$, 
\begin{align}
\sin A =& 
\frac{\sin\rho_{BC}}{\sin\rho_{AB}} , 
\label{sin-s-R}
\\
\cos A =&
\frac{\cos\rho_{BC}\sin\rho_{CA}}{\sin\rho_{AB}}
\notag\\
=&
\frac{\tan\rho_{CA}}{\tan\rho_{AB}} , 
\label{cos-s-R}
\\
\cos\rho_{AB} =&
 \cos\rho_{BC} \cos\rho_{CA} , 
 \label{cosc}
\end{align}
where 
the last equation is a generalization of Pythagorean theorem in spherical geometry. 
Eq. (\ref{cosc}) approaches $\rho_{AB}^2 = \rho_{BC}^2 + \rho_{CA}^2$ 
at very small scale of $\rho_{AB} \ll 1$.

\subsection{Gravitational lens configuration in spherical geometry}
It is usually convenient to use a correspondence between 
hyperbolic functions and spherical ones, 
when we wish to obtain expressions in spherical geometry 
from known hyperbolic ones, and vice versa. 
If this correspondence were applied to Eq. (\ref{lenseq-h-D}), 
this equation would remain the same, 
because it does not include any hyperbolic function. 
However, this is not the case as shown below. 

Figure \ref{fig-config-s} shows 
a gravitational lens system in spherical geometry. 
For the right triangle PRL, we obtain 
\begin{align}
\sin\theta 
=&
\frac{\sin \rho_{PL}}{\sin \rho_L} ,
\label{PRL-s}
\end{align}
where Eq. (\ref{sin-s-R}) is used. 
From the right triangle LQS, 
\begin{align}
\sin(\pi - \Psi_S)  
=&
\frac{\sin \rho_{QL}}{\sin \rho_{LS}} .
\label{LQS-s}
\end{align}

\begin{figure}
\includegraphics[width=8.6cm]{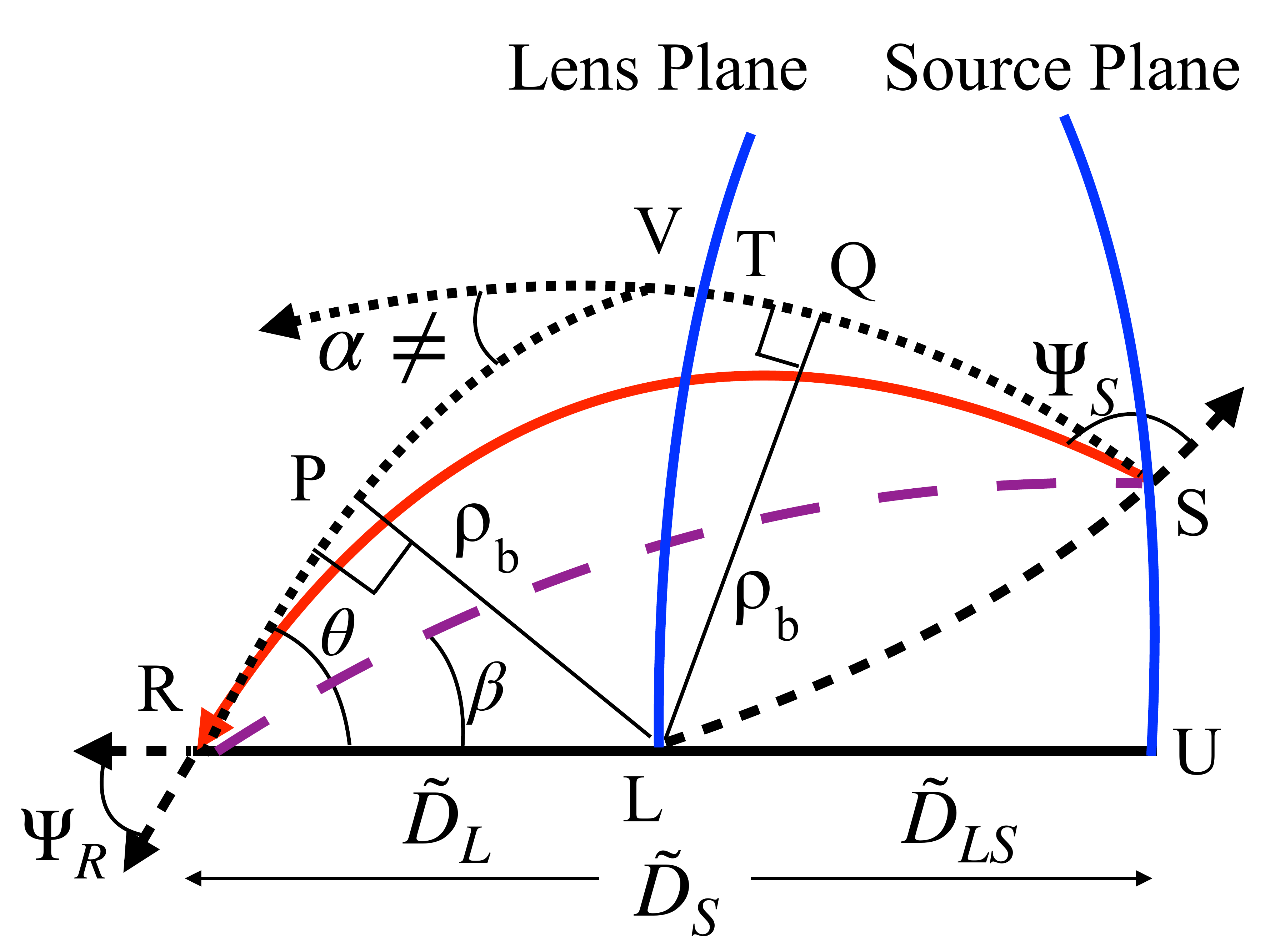}
\caption{
A gravitational lens system in AdS background. 
The notations are the same as those in Figure \ref{fig-config-h}.  
The lens and source planes are not parallel to each other, 
because they live in a curved space. 
}
\label{fig-config-s}
\end{figure}

For the single light ray, $\rho_{PL}$ and $\rho_{QL}$ 
must be equal to each other, 
because it means the impact parameter. 
Hence, $\rho_{PL} = \rho_{QL}$. 
By using this for Eqs. (\ref{PRL-s}) and (\ref{LQS-s}), 
we obtain 
\begin{align}
\Psi_S 
= 
\pi - \arcsin\left(\frac{\sin\rho_L}{\sin\rho_{LS}}\sin\theta\right) . 
\label{PsiS-s}
\end{align}

Next, we consider the right triangle LSU. 
Using Eq. (\ref{sin-s-R}) at the vertex L of LSU, 
we find 
\begin{align}
\sin(\pi - \phi_{RS}) 
=
\frac{\sin\rho_{SU}}{\sin\rho_{LS}} . 
\label{LSU-s}
\end{align}
For the triangle RSU, 
we obtain 
\begin{align}
\sin\beta 
=
\frac{\sin\rho_{SU}}{\sin\rho_S} , 
\label{RSU-s}
\end{align}
where  
$\beta$ is the angle $R$, 
namely the (unlensed) source angle in AdS spacetime. 

From Eqs. (\ref{LSU-s}) and (\ref{RSU-s}), 
a relation between $\phi_{RS}$ and $\beta$ is found as 
\begin{align}
\phi_{RS} 
= 
\pi - \arcsin\left(\frac{\sin\rho_S}{\sin\rho_{LS}}\sin\beta\right) . 
\label{phiRS-s}
\end{align}

By substituting Eqs. (\ref{PsiS-s}) and (\ref{phiRS-s}) into Eq. (\ref{alpha}), 
we obtain 
\begin{align}
\alpha - \theta 
= 
\arcsin\left(\frac{\sin\rho_L}{\sin\rho_{LS}}\sin\theta\right)
- \arcsin\left(\frac{\sin\rho_S}{\sin\rho_{LS}}\sin\beta\right) . 
\label{lenseq-AdS}
\end{align}

\subsection{Angular diameter distance in spherical geometry}
In the normalized spherical surface, 
we consider the angular diameter distance 
between two points as $\hat d$, 
while $d$ denotes 
the angular diameter distance in the original spherical space. 
Namely, 
$d = \hat d \sqrt{3/(-\Lambda)}$. 
The normalized angular diameter distances are defined as 
\begin{align}
\hat d_L 
&\equiv \sin\rho_L , 
\label{dL-s}
\\
\hat d_S 
&\equiv \sin\rho_S , 
\label{dS-s}
\\
\hat d_{LS} 
&\equiv \sin\rho_{LS} , 
\label{dLS-s}
\end{align}
such that Eq. (\ref{lenseq-AdS}) can be rewritten simply as 
\begin{align}
\alpha - \theta 
= 
\arcsin\left(\frac{\hat d_L}{\hat d_{LS}}\sin\theta\right)
- \arcsin\left(\frac{\hat d_S}{\hat d_{LS}}\sin\beta\right) . 
\label{lenseq-s-d}
\end{align}

Eq. (\ref{lenseq-s-d}) takes the same form as 
Eq. (17) of Reference \cite{Takizawa2020b} 
for the flat background. 
Note that $\hat d_L + \hat d_{LS} \neq \hat d_S$ also in spherical geometry.

\subsection{Lens and source planes in spherical geometry}
In spherical geometry, 
we consider lens and source planes 
as shown by Figure \ref{fig-config-s}. 

For the lens and source planes, 
we define 
the normalized angular diameter distance from the receiver to the lens plane 
as 
\begin{align}
\hat D_L 
&\equiv 
\hat d_L 
\notag\\
&= \sin\rho_L , 
\label{DL-s}
\end{align}
where we use Eq. (\ref{dL-s}). 
By the same way, 
the normalized angular distance from the receiver to the source plane 
is defined as 
\begin{align}
\hat D_S 
\equiv 
\sin\rho_{RU} . 
\label{DS-s}
\end{align} 
The normalized angular distance between the lens and source planes  
is the same as that between the points L and U, 
\begin{align}
\hat D_{LS} 
&\equiv 
\sin\rho_{LU} .
\label{DLS-s}
\end{align}

Note that $\rho_L + \rho_{LU} = \rho_U$ but 
$\hat D_L + \hat D_{LS} \neq \hat D_S$ in spherical geometry, 
because 
$\sin\rho_L + \sin\rho_{LU} \neq \sin\rho_U$. 
We should remember also that 
the physical angular diameter distance can be obtained 
from the normalized one by 
$D = \hat D \sqrt{3/(-\Lambda)}$.

By using the cosine formula Eq. (\ref{cos-s-R}) 
for the right triangle RSU, 
we obtain 
\begin{align}
\cos\beta 
= 
\frac{\cos \rho_{SU}\sin\rho_U}{\sin\rho_S} .
\label{cos-beta-s}
\end{align}
From Eqs. (\ref{RSU-s}) and (\ref{cos-beta-s}), 
we find 
\begin{align}
\tan\beta 
&=
\frac{\sin\beta}{\cos\beta}
\notag\\
&=
\frac{\tan\rho_{SU}}{\sin\rho_U} . 
\label{tanbeta-s}
\end{align}
We use this for Eq. (\ref{DS-s}) to obtain 
\begin{align}
\hat D_S 
= 
\frac{\tan\rho_{SU}}{\tan\beta} .
\end{align}
From this, we can see 
\begin{align}
\sin^2\rho_{SU} 
=
\frac{\hat D_S^2 \tan^2\beta}{1 - \hat D_S^2 \tan^2\beta} . 
\label{sinh2SU-s}
\end{align}

Using Eq. (\ref{cosc}) for the right triangle LSU 
leads to 
\begin{align}
\cos\rho_{LS} = \cos\rho_{SU} \cos\rho_{LU} . 
\label{cosLS}
\end{align}
This leads to 
\begin{align}
\sin^2\rho_{LS} 
&=
1 - \cos^2\rho_{LS} 
\notag\\
&=
1 - \left(\cos\rho_{LU} \cos\rho_{SU}\right)^2 
\notag\\
&=
\frac{\hat D_{LS}^2 + \hat D_S^2 \tan^2\beta}
{1 + \hat D_S^2 \tan^2\beta} ,
\label{sin2LS}
\end{align}
where Eqs. (\ref{DLS-s}) and (\ref{sinh2SU-s}) are 
used in the last line. 

By the same way to Eq. (\ref{arcsintoarctan-h}), 
we consider the angle L in the right triangle LSU to obtain 
also in spherical geometry 
\begin{align}
\arcsin
\left(
\frac{\hat d_S}{\hat d_{LS}} \sin\beta 
\right)
=
\arctan
\left(
\frac{\hat D_S}{\hat D_{LS}} \tan\beta 
\right) .
\label{arcsintoarctan-s}
\end{align}

In terms of the angular diameter distance $\hat D$, 
Eq. (\ref{lenseq-s-d}) is rearranged as 
\begin{align}
\alpha - \theta 
=& 
\arcsin\left(
\sqrt{
\frac{1 + \hat D_S^2 \tan^2\beta}{\hat D_{LS}^2 + \hat D_S^2 \tan^2\beta}
}
\hat D_L\sin\theta\right) 
\notag\\
& 
- 
\arctan\left(\frac{\hat D_S}{\hat D_{LS}}\tan\beta\right) ,
\label{lenseq-s-D}
\end{align}
where Eqs. (\ref{DL-s}), (\ref{DS-s}), (\ref{DLS-s}), (\ref{tanbeta-s}) 
(\ref{sin2LS}) and (\ref{arcsintoarctan-s}) are used. 

There exists the only difference between 
Eqs. (\ref{lenseq-h-D}) and (\ref{lenseq-s-D}). 
The sign of one term in the square root is opposite. 
For this reason, the present section avoids a conventional method of 
only replacing hyperbolic functions by spherical ones. 
The sign difference comes from  Eqs. (\ref{sinh2LS}) and (\ref{sin2LS}) 
for $\sinh\rho_{LS}^2$ and $\sin\rho_{LS}^2$, respectively. 

We should note that Eq. (\ref{lenseq-s-D}) 
is based on the angular diameter distances 
in spherical geometry due to the presence of the negative cosmological constant.

\section{Discussions}
\subsection{Unified form in Euclidean, hyperbolic and spherical geometry}
Eqs. (\ref{lenseq-h-d}) and (\ref{lenseq-s-d}) are in the same form 
as the flat lens equation. 
However, they are not in practical use, because the receiver cannot directly 
measure the distance from the lens to the source 
though the distance from the receiver to the source 
is in principle a direct observable. 
Hence, the lens equations in terms of the lens and source planes 
have much more practical use 
\cite{SEF, Petters, Dodelson, Keeton}. 

First, we unify Eqs. (\ref{lenseq-h-D}), (\ref{lenseq-f-D}) 
and (\ref{lenseq-s-D}) in a single form as 
\begin{align}
\alpha - \theta 
=& 
\arcsin\left(
\sqrt{
\frac{1 + K \hat D_S^2 \tan^2\beta}{\hat D_{LS}^2 + \hat D_S^2 \tan^2\beta}
}
\hat D_L\sin\theta\right) 
\notag\\
& 
- 
\arctan\left(\frac{\hat D_S}{\hat D_{LS}}\tan\beta\right) ,
\label{lenseq-all-D}
\end{align}
where $K$ denotes $1$, $0$ and $-1$ 
for spherical, flat and hyperbolic geometry, 
respectively. 
$K$ is corresponding to the Gaussian curvature 
of the normalized background surface. 

How does the only difference among 
Eqs. (\ref{lenseq-h-D}), (\ref{lenseq-f-D}) and (\ref{lenseq-s-D})
affect the lensed image position? 
What is the physical effect of the $K$ term in Eq. (\ref{lenseq-all-D})? 
To investigate this issue below, we shall use small angle approximations 
to follow the method by Takizawa et al. (2020b)  \cite{Takizawa2020b}.

\subsection{Iterative behaviors}
Let us introduce a book-keeping parameter $\varepsilon$. 
For a given source position angle as $\beta = \varepsilon\beta_{(1)}$, 
the image position angle and the deflection angle are expanded 
in a Taylor series as 
\begin{align}
\theta 
&= \sum_{k=1}^{\infty} \varepsilon^k \theta_{(k)} , 
\label{theta-exp}
\\
\alpha
&= \sum_{k=1}^{\infty} \varepsilon^k \alpha_{(k)} . 
\label{alpha-exp}
\end{align}

For the third-order solution in a flat background, 
please refer to Eqs. (56)-(58) of Reference \cite{Takizawa2020b} 
for instance. 
In the similar calculations to Ref. \cite{Takizawa2020b}, 
we find the third-order solution for the lens equation 
in hyperbolic/spherical geometry. 
The new term appears owing to the background curvature. 
From Eq. (\ref{lenseq-all-D}), 
we obtain the new term as 
\begin{align}
\theta_{(3)}^{New} = 
K 
\frac{\hat{D}_L\hat{D}_S^2 }{2 \hat{D}_{LS}^2 (\hat{D}_L + \hat{D}_{LS})} 
\beta_{(1)}^2 \theta_{(1)} .
\label{theta-new} 
\end{align}

At the third order level, 
$\Lambda$ ($K = -1$) decreases $\theta$ compared with in the flat background 
when the angular diameter distances are the same as each other, 
while $\theta$ is increased in AdS ($K = 1$). 
However, there is subtlety in this statement. 
Eqs. (56)-(58) of Reference \cite{Takizawa2020b} uses 
the addition law 
$D_S = D_L + D_{LS}$ because of the flat background. 
In a curved background, however, they have to be modified 
because $\hat{D}_L + \hat{D}_{LS} \neq \hat{D}_S$. 

The totally modified form is thus 
\begin{align}
\theta_{(3)} 
=& 
\frac{\hat{D}_{LS}}{\hat{D}_L + \hat{D}_{LS}} 
\alpha_{(3)} 
-  
\frac{\hat{D}_S}{3 (\hat{D}_L + \hat{D}_{LS})}
\left(
1 - \frac{\hat{D}_S^2}{\hat{D}_{LS}^2} 
\right) 
\beta_{(1)}^3
\notag\\
&
- \frac{(1-K) \hat{D}_L\hat{D}_S^2 }{2 \hat{D}_{LS}^2(\hat{D}_L + \hat{D}_{LS})} 
\beta_{(1)}^2 \theta_{(1)} 
\notag\\
&
-  
\frac{\hat{D}_L}{6(\hat{D}_L + \hat{D}_{LS})}
\left(
1 - \frac{\hat{D}_L^2}{\hat{D}_{LS}^2} 
\right) 
\theta_{(1)}^3 ,
\label{theta-3}
\end{align}
where the new term makes a correction in the second line. 
Instead of assuming a specific model of the lens object, 
here, we consider a general one, for which the deflection angle of light 
at $O(\varepsilon^3)$ is denoted as $\alpha_{(3)}$. 
$\alpha_{(3)}$ can be calculated by using the lower order solutions 
$\theta_{(1)}$ and $\theta_{(2)}$ for a given lens model. 
In some model, $\theta_{(2)}$ vanishes \cite{Takizawa2020b}. 

To be rigorous, $\theta$ is not always a direct observable, 
because it is the angle measured from the lens direction 
but the direction is unknown in several cases. 
On the other hand, 
the separation angle between two lensed images, 
each of which is located on the opposite sides of the lens, 
is a direct observable. 
The separation angle is decreased (increased) by 
not $\Lambda > 0$ ($\Lambda < 0$) alone 
but its coupling with the lens mass. 

The above correction due to the new $K$ term is 
a mathematical consequence of introducing the lens and source planes 
into curved backgrounds.

\subsection{Deflection angle of light on dS/AdS backgrounds}
Up to this point, we have not mentioned details of $\alpha$. 
Takizawa et al. have shown Eq. ($\ref{alpha}$) for the deflection angle of light 
can be justified even in a non-asymptotically flat spacetime \cite{Takizawa2020a}. 
They have calculated $\alpha$ in the Schwarzschild de-Sitter spacetime. 
For their $\alpha$, the background spacetime is implicitly Minkowskian, 
because $\alpha$ vanishes only when $M = 0$ and $\Lambda = 0$. 
Hence, their $\alpha$ should be used 
in Eq. (\ref{lenseq-f-D}). 

On the other hand, 
Eqs. (\ref{lenseq-h-d}) and (\ref{lenseq-h-D}) are 
valid in hyperbolic geometry due to the positive cosmological constant, 
while 
Eqs. (\ref{lenseq-s-d}) and (\ref{lenseq-s-D}) hold 
in spherical geometry due to the negative cosmological constant. 

In the present method, 
the effects of purely the cosmological constant are included 
in the angular diameter distance of the dS/AdS backgrounds.  
Therefore, the deflection angle of light $\alpha^{dS}$ is subtracted by 
the effects of purely $\Lambda$. 
For the clarity, we denote the deflection angle of light explicitly as 
$\alpha^{dS}(p_i, \Lambda)$, 
where the lens object is parameterized by 
$p_i$  ($i = 1, 2, \cdots $) in addition to $\Lambda$. 
In the Kerr de-Sitter spacetime for instance, 
$p_i$ corresponds to the mass or spin parameter.  
The deflection angle of light on dS/AdS backgrounds is thus  
\begin{align}
\alpha^{dS}  
\equiv \alpha(p_i, \Lambda) - \alpha(p_i=0, \Lambda) .
\label{delta-alpha}
\end{align}

\begin{figure}
\includegraphics[width=8.6cm]{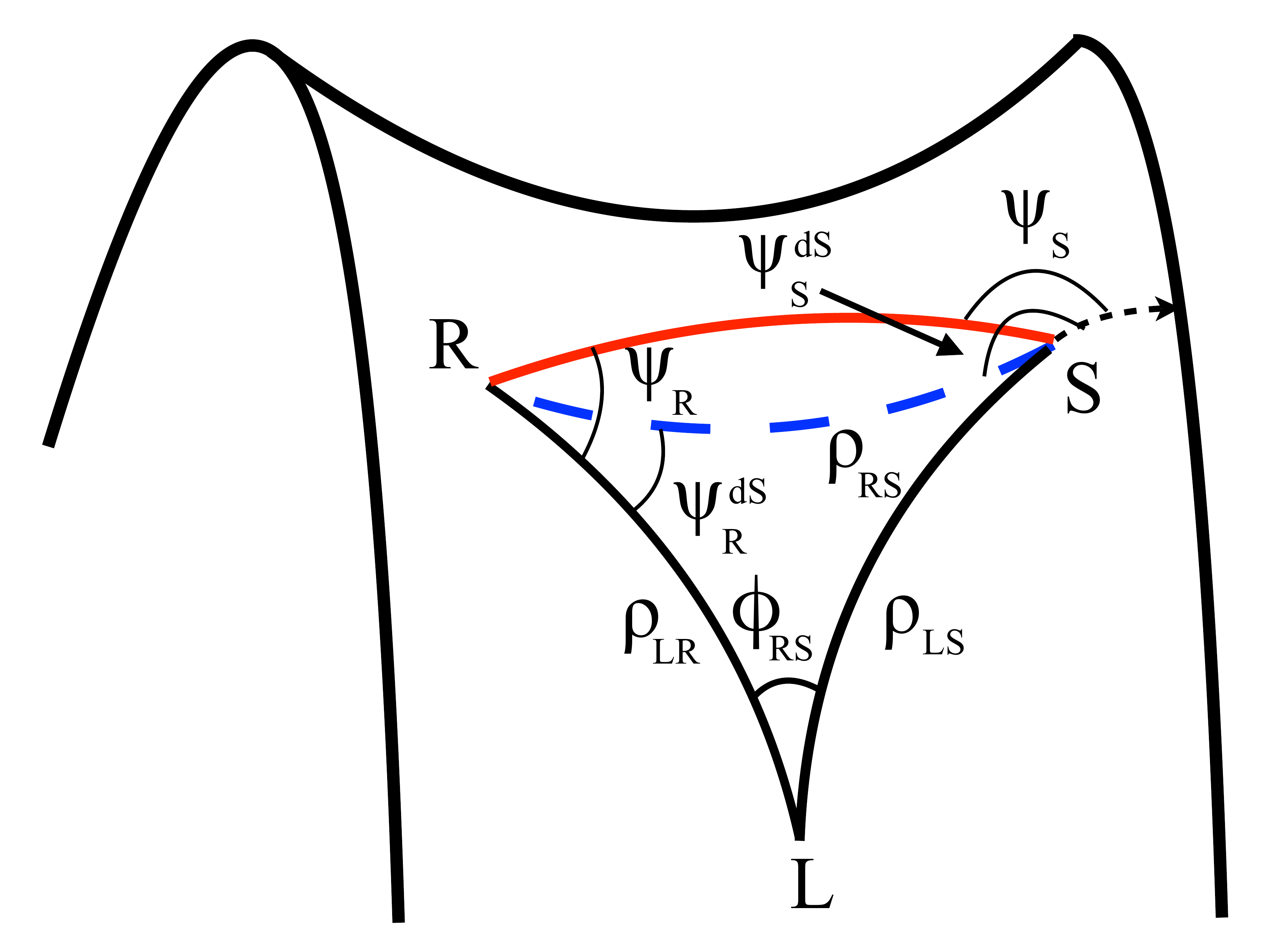}
\caption{
Schematic figure of the deflection angle of light on a hyperbolic plane 
as the dS background. 
The solid lines LR and LS denote the radial geodesics 
emanating from the lens to the receiver and the source, respectively. 
The inner angle at L is $\phi_{RS}$ which is a given parameter 
characterizing the angle separation between R and S. 
The dashed blue (in color) curve denotes 
the geodesic connecting S and R.  
The solid red (in color) curve denotes 
a {\it true} light ray from S to R. 
The light ray by a gravitating mass in the presence of $\Lambda$ 
does not follow a geodesic in hyperbolic geometry. 
The angle between the radial line LR and the geodesic RS is denoted as 
$\Psi_R^{dS}$, 
while that between the radial line LR and the light ray 
is denoted as $\Psi_R$. 
The dotted arrow denotes 
the tangent of the radial line LS. 
The angle between the radial line LS and the geodesic RS 
is an outer angle, which is denoted as $\Psi_S^{dS}$. 
The angle between the radial line LS and the light ray 
is an outer angle, which is denoted as $\Psi_S$. 
We consider the tangents of the solid red (in color) line 
and those of the dashed blue (in color) one 
at the two points R and S. 
The angle difference between the two tangents 
at the source position S is 
$\Psi_S - \Psi_S^{dS}$, 
which is denoted as $\delta_S$.  
The angle difference between the two tangents 
at the receiver point R is 
$\Psi_R^{dS} - \Psi_R$, 
which is denoted as $\delta_R$. 
The right-had side of Eq. (\ref{delta-alpha-2}) 
is the sum of the two angle differences, namely 
$\delta_R + \delta_S$. 
This can be thus interpreted as the deflection angle of the solid red (in color) line 
to the reference line as the dashed blue (in color) one. 
}
\label{fig-deflection-h}
\end{figure}

Let us explain why Eq. (\ref{delta-alpha}) is justified. 
See Figure \ref{fig-deflection-h} for two triangles LRS in a hyperbolic plane. 
One triangle LRS has a side RS that is a hyperbolic geodesic indicated by 
a dashed blue (in color) line, 
while the other LRS has another side RS that means a {\it true} light ray 
denoted by a solid red (in color) line. 
From Figure \ref{fig-deflection-h}, we find 
$\alpha(p_i, \Lambda) = \Psi_R - \Psi_S + \phi_{RS}$, 
and 
$\alpha(p_i, = 0, \Lambda) = \Psi_R^{dS} - \Psi_S^{dS} + \phi_{RS}$. 
By using Eq. (\ref{delta-alpha}), therefore, we obtain 
\begin{align}
\alpha^{dS}  
= 
(\Psi_R - \Psi_R^{dS}) + (\Psi_S^{dS} - \Psi_S) .
\label{delta-alpha-2}
\end{align}

This allows us to interpret $\alpha^{dS}$
as the deflection angle of the light ray (the solid red (in color) line) 
with respect to the reference line (the dashed blue (in color) line). 
This point is explained also in the caption of Figure \ref{fig-deflection-h}. 
As a result, Eq. (\ref{delta-alpha}) has the meaning of the deflection angle 
of light on the dS background 
\cite{footnote-4}. 
By the same way, one can see that Eq. (\ref{delta-alpha}) 
gives the deflection angle of light on the AdS background 
as the spherical surface.

As an example, 
we assume the Schwarzschild de-Sitter spacetime, 
for which $\alpha$ is calculated in References 
\cite{Ishihara2016, Takizawa2020a}. 
Their $\alpha$ is invariant 
under transformations of spatial coordinates, 
because it can be expressed as the areal integral of the Gaussian 
curvature of the plane. 
By using Eq. (\ref{delta-alpha}) for their expression of $\alpha$, 
we obtain the deflection angle 
on the dS/AdS backgrounds as
\begin{align}
\alpha^{dS}
=&
  \frac{r_g}{b}\left(\sqrt{1 - b^2 u^2_S} + \sqrt{1 - b^2 u^2_R}\right)
\notag\\
& + \frac{r_g b\Lambda}{12}\left(\frac{1}{\sqrt{1 - b^2 u^2_S}}
  + \frac{1}{\sqrt{1 - b^2 u^2_R}}\right)
\notag\\
&  + O\left(r^2_g, \Lambda^2\right) , 
\label{delta-alpha-dS}
\end{align}
where $r_g \equiv 2m$ for the mass $m$. 

Note that, in the present formulation,  
$\alpha^{dS}$ does not include any term of purely the cosmological constant. 
This is because we work on the dS background and hence 
the effects of purely the cosmological constant are fully included 
in the well-defined angular diameter distance. 

On the dS background, 
the impact parameter of light $b$ is 
$b = \sqrt{3/\Lambda} R_b = \sqrt{3/\Lambda} \tanh \rho_b$, 
where 
$R_b$ denotes the normalized impact parameter 
and 
$\rho_b$ is the arc length corresponding to 
the impact parameter of light. 
\begin{align}
\frac{r_g}{b} 
=&
\sqrt{\frac{\Lambda}{3}} \frac{r_g}{\tanh\rho_b} 
\notag\\
=&
r_g \sqrt{\frac{\Lambda}{3}} 
\left(
\frac{1}{\hat D_b} + \frac{\hat D_b}{2} + O(\hat D^2_b) 
\right)
\notag\\
=&
\frac{r_g}{D_b} 
+ 
\frac{r_g \Lambda D_b}{6} 
+O\left( r_g \Lambda^2 D_b^3 \right) ,
\label{2m-1}
\end{align}
where we use in the second line 
\begin{align}
\frac{1}{\tanh\rho_b} 
= 
\frac{1}{\sinh\rho_b} + \frac{\sinh\rho_b}{2} 
+ O\left(\sinh^3\rho_b\right) , 
\label{tanh-inverse}
\end{align}
and 
the angular diameter distance and the normalized one 
corresponding to $\rho_b$ 
are denote denoted as 
$\hat D_b \equiv \sinh\rho_b$
and 
$D_b \equiv \sqrt{3/\Lambda} \hat D_b$, 
respectively.  

By using Eq. (\ref{sin-h-R}) for the right triangle LPR 
in Figure \ref{fig-config-h}, 
we obtain
\begin{align}
\sin\theta = \frac{D_b}{D_L} . 
\end{align}
By using this for $D_b$ in Eq. (\ref{2m-1}), 
we find 
\begin{align}
\frac{r_g}{b} 
=&
\frac{r_g}{D_L \sin\theta} 
+ 
\frac{r_g \Lambda D_L \sin\theta}{6} 
+O\left( r_g \Lambda^2 D_b^3 \right) . 
\label{2m-2}
\end{align}

For more simplicity, 
we employ small angle approximations. 
Then, we find 
\begin{align}
b^2u^2_R 
&= 
\theta^2 
+ \frac13 \Lambda D_L^2 \theta^2 
+ O(\theta^4, \theta^4 D^2 \Lambda, \theta D^4 \Lambda^2) , 
\label{buR}
\\
b^2u^2_S 
&= 
\left( \frac{D_L}{D_{LS}} \right)^2 \theta^2 
+ \frac13 \Lambda D_L^2 \theta^2 
+ O(\theta^4, \theta^4 D^2 \Lambda, \theta D^4 \Lambda^2) , 
\label{buS}
\end{align}
where 
the latter equation can be obtained by noting Eq. (\ref{PsiS-h}). 
By using Eqs. (\ref{2m-2}), (\ref{buR}) and (\ref{buS}), 
Eq. (\ref{delta-alpha-dS}) is simplified as 
\begin{align}
\alpha^{dS}
=&
  \frac{2 r_g}{D_L \theta} 
- \frac{r_g \theta}{2D_L} 
\left[  
1 + \left(\frac{D_L}{D_{LS}}\right)^2 
\right]
\notag\\
&
  + \frac{r_g \Lambda D_L \theta}{6} 
 + O\left(r_g^2, r_g \theta^3, r_g \Lambda D \theta^3, r_g \Lambda^2 D^3 \right) .
\label{delta-alpha-dS-2}
\end{align}

In the AdS case, $\tanh$ functions 
should be replaced by $\tan$ ones. 
This leads to 
\begin{align}
\frac{1}{\tan\rho_b} 
= 
\frac{1}{\sin\rho_b} - \frac{\sin\rho_b}{2} 
+ O\left(\sin^3\rho_b\right) , 
\label{tan-inverse}
\end{align}
which has a difference in the sign of 
the second term of the right-hand side, 
compared with Eq. (\ref{tanh-inverse}). 
However, we obtain 
\begin{align}
\frac{r_g}{b} 
=&
\frac{r_g}{D_L \sin\theta} 
+
\frac{r_g \Lambda D_L \sin\theta}{6} 
+O\left( r_g \Lambda^2 D_b^3 \right) ,
\label{2m-AdS}
\end{align}
where the second term of the right-hand side has 
the same sign as the dS case 
because of the factor $\sqrt{3/(-\Lambda)}$ 
in the angular diameter distance, 
though 
the second term in the right-hand side of Eq. (\ref{tan-inverse}) 
has the minus sign.  
As a consequence, 
we obtain for the AdS 
\begin{align}
\alpha^{AdS}
=&
  \frac{2 r_g}{D_L \theta} 
- \frac{r_g \theta}{2D_L} 
\left[  
1 + \left(\frac{D_L}{D_{LS}}\right)^2 
\right]
\notag\\
&
  + \frac{r_g \Lambda D_L \theta}{6} 
 + O\left(r_g^2, r_g \theta^3, r_g \Lambda D \theta^3, r_g \Lambda^2 D^3 \right) .\label{delta-alpha-AdS}
\end{align}

Eq. (\ref{delta-alpha-AdS}) is in the same form as 
Eq. (\ref{delta-alpha-dS-2}), 
though the two background geometries are very different.

In order to discuss effects of $\Lambda$ 
on lensing observations, we consider 
Eqs. (\ref{lenseq-h-d}) and (\ref{lenseq-s-d}).  
In terms of the angular diameter distances 
among the three points L, R and S,  
the dS/AdS lens equations take exactly the same form as the flat one. 
We assume an ideal situation that 
the angular diameter distances 
$d_L$, $d_S$ and $d_{LS}$ take the same values 
for each of the flat, dS and AdS cases. 
Then, the difference in the three lens equations can come only from 
the deflection angle of light. 
According to Eqs. (\ref{delta-alpha-dS-2}) and (\ref{delta-alpha-AdS}), 
the deflection angle of light for a given lens mass 
is larger (smaller) in dS (AdS) than in the flat case, 
where $d_L = D_L$ is used. 
This means that, 
through a coupling of $\Lambda$ with $m$, 
the separation angle of multiple images 
is increased (decreased) by $\Lambda > 0$ ($\Lambda < 0$), 
for given $m$, $\beta$, $d_L$, $d_S$ and $d_{LS}$.

\begin{figure}
\includegraphics[width=8.0cm]{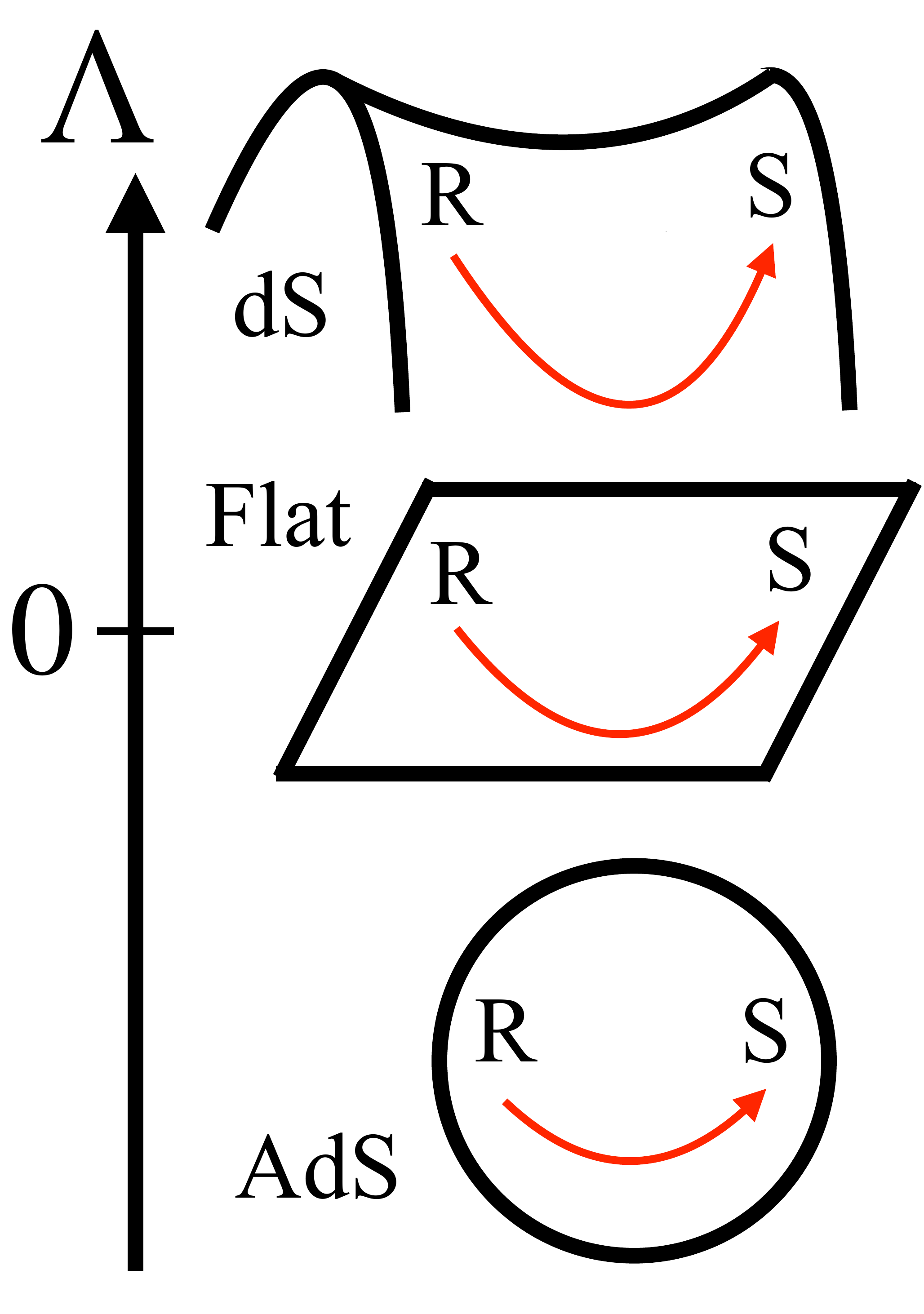}
\caption{
Lensed light rays on three distinct manifolds 
that are characterized by $\Lambda$. 
Each light ray is denoted by a solid red (in color) arrow. 
}
\label{fig-topology}
\end{figure}

It is worthwhile to mention also that 
$\alpha$ in References \cite{Ishihara2016, Takizawa2020a} 
apparently diverges as $b \to \infty$, 
while Eq. (\ref{delta-alpha-dS}) is not divergent 
because the present formulation takes full account of 
the curvature of dS backgrounds. 
This means that the present formulation can describe 
the lensing behavior on much larger scale than 
the conventional method based on the Minkowskian background. 
See Figure \ref{fig-topology} for a schematic illustration of 
light rays on hyperbolic, flat and spherical surfaces. 

\subsection{Strong deflection}
Finally, we mention the strong deflection case, 
for which light rays have the winding number $N$, 
where $N$ is a positive integer. 
We thus extend Eq. (\ref{lenseq-all-D}) as 
\begin{align}
\alpha - \theta 
=& 
\arcsin\left(
\sqrt{
\frac{1 + K \hat D_S^2 \tan^2\beta}{\hat D_{LS}^2 + \hat D_S^2 \tan^2\beta}
}
\hat D_L\sin\theta\right) 
\notag\\
& 
- 
\arctan\left(\frac{\hat D_S}{\hat D_{LS}}\tan\beta\right) 
+ 2\pi N .
\label{lenseq-strong-D}
\end{align}
Fully nonlinear investigations of the strong deflection 
with Eq. (\ref{lenseq-strong-D}) are left for future.

\section{Summary}
Gravitational lens equations have been discussed in dS background, 
for which the existence of the dS horizon has been taken into account and 
hyperbolic trigonometry has been used together with 
the hyperbolic angular diameter distance. 
We have used spherical trigonometry in order to discuss 
gravitational lens equations in AdS background. 

In terms of the angular diameter distances between two points, 
the lens equations on the dS/AdS backgrounds 
Eqs. (\ref{lenseq-h-d}) and (\ref{lenseq-s-d})
take exactly the same form as that in the flat background 
\cite{Takizawa2020b}. 
On the other hand, 
there exists a difference among those 
in terms of the angular diameter distances 
using the lens and source planes. 
See Eqs. (\ref{lenseq-h-D}), (\ref{lenseq-f-D}) and (\ref{lenseq-s-D}). 
The only difference in the form is indicated as 
the $K$ term in Eq. (\ref{lenseq-all-D}). 

In small angle approximations, 
the difference in the form among the dS/AdS lens equations 
and the exact lens equation in Minkowski background 
begins at the third order. 
The angular separation of lensed images is 
decreased by the third-order deviation in the dS lens equation, 
while it is increased in AdS. 

We have discussed also the deflection angle of light to match with 
the lens equations on the dS/AdS backgrounds. 
This new form of the deflection angle of light 
does not include any term of purely the cosmological constant.  
In addition, 
we wish to stress that 
the deflection angle of light rays 
in both hyperbolic and spherical geometry 
can take the same form within the present framework. 
It would be worthwhile to fully understand a reason for the coincidence. 

Through a coupling of $\Lambda$ with $m$, 
the separation angle of multiple images 
is increased (decreased) by $\Lambda > 0$ ($\Lambda < 0$), 
for a given mass, source direction and angular diameter distances 
among the lens, receiver and source.

The above results imply that 
a similar behavior in the dS/AdS lensing may occur 
at the third order level 
of the general relativistic cosmological perturbations. 
Along this direction, a study on realistic cosmological backgrounds 
with $\Lambda$ is left for future.

\begin{acknowledgments}
We thank Marcus Werner and Toshiaki Ono for the useful discussions 
on the optical metric and the background subtraction. 
We wish to thank Mareki Honma for the conversations 
on the EHT method and technology. 
We thank Hideyoshi Arakida, Yuuiti Sendouda, Ryuichi Takahashi, 
Masumi Kasai,  Ryuya Kudo, 
Ryunosuke Kotaki, Masashi Shinoda, and  Hideaki Suzuki 
for the useful conversations. 
This work was supported 
in part by Japan Society for the Promotion of Science (JSPS) 
Grant-in-Aid for Scientific Research, 
No. 20K03963 (H.A.), 
and 
in part by Ministry of Education, Culture, Sports, Science, and Technology, 
Grant No. 17H06359 (H. A.) 
\end{acknowledgments}


\begin{thebibliography}{99}
\bibitem{footnote-1}
Instead of $\rho$, if one introduces $\chi$ coordinate by $R \equiv \sin\chi$ 
as usually done in cosmology, 
the proper length $L$ on the constant time slice is expressed as 
$dL^2 =  d\chi^2 + \sin\chi^2( d\Theta^2 + \sin^2 \Theta d\phi^2)$, 
with which cosmologists are familiar. 
The proper length $L$ describes a positive constant curvature space, 
namely spherical geometry. 
However, the proper length does not matter 
in the light propagation.
\bibitem{footnote-2}
The proper length in the AdS spacetime is known to describe 
a negative constant curvature space, namely hyperbolic geometry. 
The present paper does not make any use of the proper length 
for discussing the light propagation.
\bibitem{footnote-4}
In the present paper, 
curves on two different manifolds are not simultaneously computed 
for a comparison. 
The calculations are done separately for dS and for AdS. 
This is because an exact method of precisely comparing curves 
in different manifolds has not been established so far.
\end{thebibliography}
\end{document}